\documentclass{PoS}

\usepackage{amsmath}
\usepackage{amsfonts}
\usepackage{braket}
\usepackage{float}
\usepackage{pgf}
\usepackage{mathtools}
\usepackage{dsfont}
\usepackage{subcaption}
\newcommand{\rw}{\rho_{\mu \nu}(\omega)}

\title{Transport coefficients of the QGP}

\ShortTitle{Transport coefficients of the QGP}

\author{\speaker{Alessandro Amato}\\
Department of Physics, College of Science, 
  Swansea University, Swansea, United Kingdom\\
  Institut f\"ur theoretische Physik, Universit\"at Regensburg, 
  Regensburg, Germany\\
  E-mail: \email{Alessandro.Amato@physik.uni-regensburg.de}}

\author{Gert Aarts\\
  Department of Physics, College of Science, 
  Swansea University, Swansea, United Kingdom\\
  E-mail: \email{g.aarts@swan.ac.uk}}

\author{Chris Allton\\
  Department of Physics, College of Science, 
  Swansea University, Swansea, United Kingdom\\
  E-mail: \email{c.allton@swan.ac.uk}}
\author{Pietro Giudice\\
  Universit\"at M\"unster, Institut f\"ur Theoretische Physik,
  M\"unster, Germany\\
  E-mail: \email{p.giudice@uni-muenster.de}}

\author{Simon Hands\\
  Department of Physics, College of Science, 
  Swansea University, Swansea, United Kingdom\\
  E-mail: \email{s.hands@swansea.ac.uk}}

\author{Jon-Ivar Skullerud\\
  Department of Mathematical Physics, National University of Ireland Maynooth, 
  Maynooth, County Kildare, Ireland\\
  E-mail: \email{jonivar@thphys.nuim.ie}}

\abstract{

The {\sc fastsum} collaboration presents a study on
the temperature dependence of the electrical conductivity
$\sigma$ in the quark-gluon plasma, using the methods of lattice QCD.
Correlators of the exactly conserved vector current are measured at different
temperatures across the deconfinement transition, using ensembles of $2+1$
flavours of dynamical fermions on anisotropic lattices.  
We then employ bayesian methods (MEM) to extract the relevant
spectral functions, which are found to be consistent with $\sigma/T$ rising as a
function of $T$.
The robustness of the results is verified by a detailed analysis of
the systematics involved in the bayesian reconstruction of the spectral
functions.}

\FullConference{31st International Symposium on Lattice Field Theory - LATTICE
2013\\
		July 29 - August 3, 2013\\
		Mainz, Germany}

\begin{document}
\section{Introduction}

In this work we study the dynamical properties of the quark-gluon plasma (QGP),
and
in particular we focus on the transport phenomena associated with electric charge \cite{Amato:2013naa}. 
From an experimental point of view, a
reliable first-principles determination of the conductivity is necessary to
estimate the evolution of electromagnetic fields during heavy-ion collision experiments, where particle emission from the fireball is heavily
influenced by
the presence of a magnetic field \cite{Tuchin:2013ie}.
In order to understand the real-time evolution of the plasma, effective
models, such as dissipative hydrodynamics, are used
\cite{effective}. Transport coefficients, e.g.\ the
bulk/shear viscosity and the electrical conductivity, enter as parameters in
these models
and can be seen as low-energy variables describing the effects of the underlying
quantum field theory. 
There are many phenomenological indications
that the plasma produced in such collisions is strongly
interacting. In many cases this leads to a very poor convergence
of
perturbation theory when used as a tool to derive expressions for transport coefficients
directly from QCD, see e.g.\ results for the shear viscosity
\cite{Arnold:2000dr}. This is why we take here the lattice QCD approach, where these quantities may be extracted from non-perturbatively computed euclidean correlators.
For recent reviews, see e.g.\ Ref.~\cite{review}.

\section{Electrical Conductivity on the Lattice}
In the following we will describe how to formulate the problem of calculating
the electrical conductivity  on the lattice. 
As a starting point, we consider the contribution from the up and
down quarks to the electromagnetic current:
\begin{equation}\label{emcur}
j^{\rm em}_\mu(x) =
\frac{2}{3}e\,j^{\;\text{up}}_\mu(x) -\frac{1}{3}e\,j^{\;\text{down}}_\mu(x), 
\end{equation} 
 where  $j_\mu^f$ is the vector current for each flavour considered, the
coefficients in front represent the fractional charge of the quark and $e$ is the elementary charge.
The connection between the conductivity and lattice QCD is provided by
the spectral function $\rw$, which is defined through the euclidean
correlator of the electromagnetic current (\ref{emcur}) as follows:
\begin{equation}
 G^{\;\text{em}}_{\mu\nu}(\tau) = \int d^3x\, 
 \braket{\,j^{\;\rm em}_\mu(\tau,{\mathbf x}) j^{{\; \rm em}}_\nu(0,{\mathbf 0})^\dagger\,} 
 =  \int_0^\infty \frac{d\omega}{2\pi}\,
 \rho^{\rm
em}_{\mu\nu}(\omega)\,
 \frac{\cosh[\omega(\tau-1/2T)]}{\sinh[\omega/2T]}
, \label{spectr} 
\end{equation}
where $\tau$ is the euclidean time coordinate, $\omega$ the frequency and 
we performed the projection to zero momentum.
On the r.h.s. the integral transform has a nontrivial kernel $K(\tau,\omega)$,
which depends on the temperature $T$.
The operation of obtaining the spectral function $\rw$ from the
euclidean correlator $G_{\mu\nu}^{\,\text{em}}(\tau)$, i.e.\ inverting
Eq.\ (\ref{spectr}),
represents the main computational challenge as it
involves analytical continuation from euclidean to Minkowski  space. Our
approach
is described in Sec.~\ref{sec:mem}. 
Once the spectral function is available, the electrical conductivity $\sigma$
can be obtained by taking the low-frequency limit, known as Kubo's formula
\cite{Kadanoff1963419}:
\begin{equation}\label{kubo}
  \frac{\sigma}{T} = \frac{1}{6T}  \lim_{\omega \rightarrow 0}\frac{\rho^{\rm
em}(\omega)}{\omega}, 
 \quad\quad
 \rho^{\rm em}(\omega) = \sum_{i=1}^3 \rho^{\rm em}_{ii}(\omega)\;.
\end{equation}

\section{Simulation Details}\label{sec:simul}

\begin{table}[!t]
\begin{subtable}{.6\linewidth}
      \begin{center}
\begin{tabular}{ccccccc}
    \hline\hline
    \ $N_s$\ &\ $N_\tau$\  &\ $T$\ [MeV] &\ $T/T_c$\  &\ $N_{\texttt{CFG}}$\ &\
$N_{\texttt{SRC}}$\  \\
\hline\hline
   32 & 16 & 352 & 1.90 & 1059  &4 \\ 
   24 & 20 & 281 & 1.52 & 1001  &4 \\ 
   32 & 24 & 235 & 1.27 & 500   &4 \\ 
   32 & 28 & 201 & 1.09 & 502   &4 \\ 
   32 & 32 & 176 & 0.95 & 501   &4 \\ 
   24 & 36 & 156 & 0.84 & 501   &4 \\
   24 & 40 & 141 & 0.76 & 523   &4 \\
   32 & 48 & 117 & 0.63 & 601   &1 \\
   \hline\hline
   \end{tabular}
\caption{Gauge configurations.}\label{tab:lattice-a}
\end{center}
    \end{subtable}%
    \begin{subtable}{.4\linewidth}      
   \begin{center}
\begin{tabular}{rl}
       \hline\hline
   $a_s$\ [fm] & $0.1227(8)$  \\
   $a_\tau$\ [fm] &  $0.03506(23)$ \\
   $\xi=a_s/a_\tau$ &$3.5$ \\
   $\gamma_g$ &$4.3$ \\
   $\gamma_f$ &$3.4$ \\
   $c_t$ & $0.9027$ \\
    $c_s$ & $1.5893$ \\
   $\hat{m}_{\rm u,d}$ &$ -0.0840$ \\
$\hat{m}_{\rm s}$&$ - 0.0743$ \vspace{0.08cm}\\
   \hline\hline
    \end{tabular}
\caption{Simulation parameters.}\label{tab:lattice-b}
    \end{center}
    \end{subtable} 
\caption{ $(a)$ The gauge ensembles have lattice size of
$N_s^3\times N_\tau$, with $N_{\texttt{CFG}}$ configurations available for each
set
and a number of $N_{\texttt{SRC}}$ sources for the analysis.
$(b)$ The parameters in the action (\protect\ref{clover}). }
\label{tab:lattice}
\end{table}

In contrast to lattice simulations used to study QCD thermodynamics, 
 where the staggered formulation is often preferred,  we take here the choice of
clover-improved Wilson fermions, with $2+1$ flavors
\cite{anis}. This is motivated by the fact that matching the physical degrees of freedom is
much easier. In fact, with staggered fermions, the euclidean correlator
described
above, which represents our main probe of the QGP, contains
a signal from an opposite parity partner, 
effectively reducing the number of usable points in the temporal direction
\cite{Aarts:2007wj}. This is undesired, since a higher resolution in $G_{\mu\nu}^{\,\text{em}}(\tau)$ brings a more reliable analysis. 
These requirements, together with the need to
keep the computational cost under control, have motivated us to simulate 
using anisotropic lattices, where  the lattice
spacing $a_\tau$ in the time direction is taken to be smaller than the spatial
one $a_s$.
The drawback of this choice is the appearance of new bare
parameters  in the action, which have to be tuned carefully. This has been achieved in Ref.\ 
\cite{anis}, to which we refer for further details. The gauge action is  
Symanzik and tadpole improved with tree-level coefficients. 
The Dirac operator reads:
\begin{equation}
 \label{clover}
 D[U] =
  \hat{m}_0 +  \gamma_4 \hat{W}_4 +\frac{1}{\gamma_f}  \sum_i \gamma_i \hat{W}_i
  - \frac{c_t}{2} \sum_{i}\sigma_{4i}\hat{F}_{4i} 
 -  \frac{c_s}{2\gamma_g}\sum_{i<j} \sigma_{ij} \hat{F}_{ij},
\end{equation}
where $\hat{m}_0$ and $\hat W_\mu$ are the mass and the usual Wilson operator
and $\gamma_\mu$ are the Dirac matrices. The last two terms are the
clover operators, with $\sigma_{\mu\nu}  =  \frac{i}{2}[\gamma_{\mu},
\gamma_{\nu}]$ and
$\hat{F}_{\mu\nu}$ the lattice version of the field strength tensor. Their
coefficients $c_{t,s}$ have been chosen according to tree-level conditions. 
The novel parameters mentioned above are the bare gauge ($\gamma_g$)  and fermion
($\gamma_f$) anisotropies. These are tuned in Ref.\ \cite{anis}, giving a ratio $
\xi \equiv a_s / a_\tau = 3.5$. 
The gauge links $U_\mu$ are represented by three-dimensional stout-smeared links
\cite{smear}, with smearing weight $\rho=0.14$ and $n_\rho=2$ iterations.
The light and strange quark mass parameters are chosen  \cite{anis} to
reproduce the physical strange quark mass and a light quark mass with
$M_\pi/M_\rho = 0.446(3)$. 
The numerical value of all the parameters appearing in Eq.\ (\ref{clover}) can be
found in Table \ref{tab:lattice-b}.

We have generated a number of
finite-temperature ensembles, using a fixed lattice spacing approach. 
This
allows for a determination of the temperature dependence of the conductivity,
missing in previous studies.
The critical temperature is estimated from the
renormalized Polyakov loop
inflection point \cite{inprep1}, see Table \ref{tab:lattice-a}. 
\begin{figure}[t]
    \centering
\includegraphics[width=\textwidth]{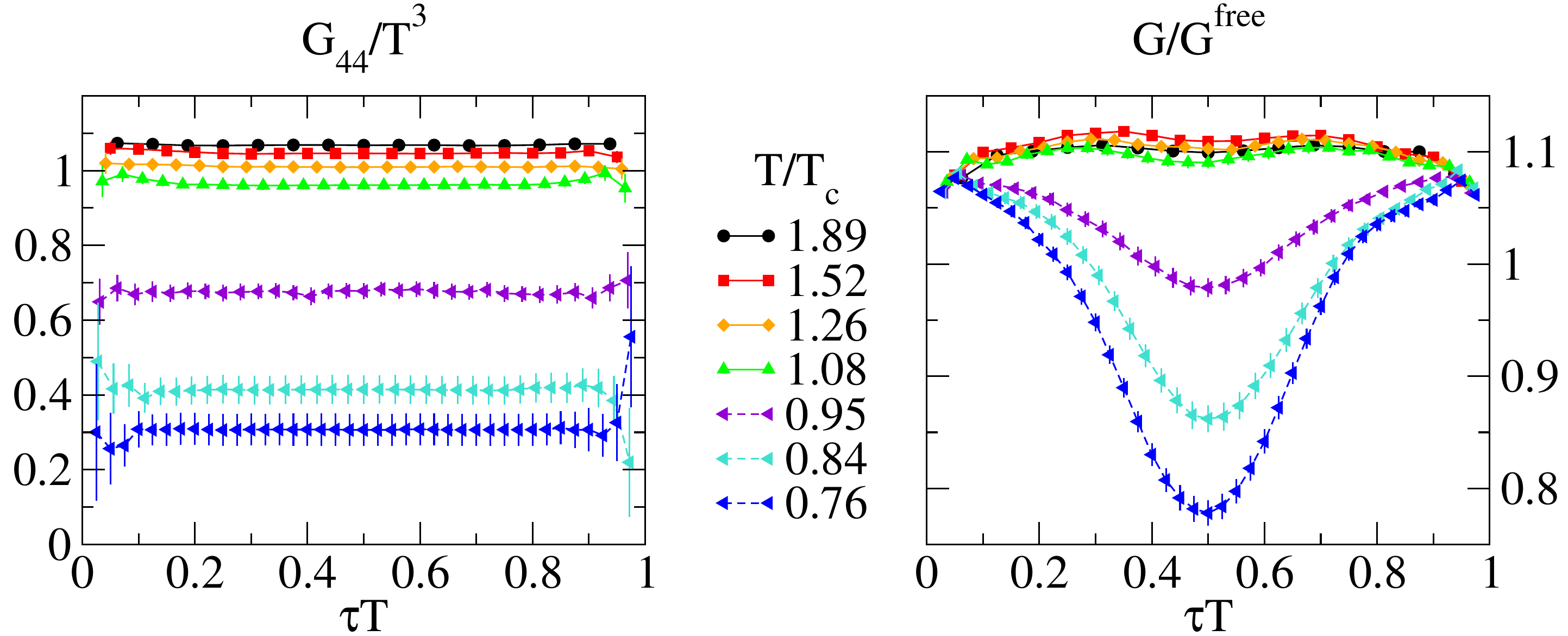}
\caption{ Diagonal components of the correlator (\protect\ref{spectr}).
Dashed (solid) lines represent temperatures below (above) $T_c$.
Left: temporal component $G_{44}$ rescaled by $T^3$.
Right: spatial component $G(\tau)=\sum_i G_{ii}(\tau)$, normalized by its value
in the free case.
}
\label{fig:corr}
\end{figure}
We use the exactly
conserved vector current on the lattice:
\begin{equation}
V_{\mu}^{\texttt{C}}(x) = \kappa_\mu \bigg[  \bar\psi(x+\hat\mu)(1+\gamma_\mu) \,
U_\mu^\dagger(x) \, \psi(x)
  - \bar\psi(x) (1-\gamma_\mu) \, U_\mu(x) \, \psi(x+\hat\mu)\bigg],  
\end{equation}
where $\kappa_4=1/2$, $\kappa_i=1/(2\gamma_f)$.
To compute the correlator (\ref{spectr}), we
use Wick contractions and neglect disconnected diagrams. This is justified by
the fact that their contribution is identically zero in the $N_f=3$ case, and we
note that the same choice has been applied in all previous studies
\cite{Aarts:2007wj,gupta,fit}.
Since the up and down quark fields in Eq.\ (\ref{emcur}) are
degenerate, we can factor out their fractional charge
assignments, via $C_{\rm em} = e^2( q_u^2 + q_d^2)= 5/9e^2$ and define 
$G^{\rm em}_{\mu\nu}(\tau) = C_{\,\rm em}\,G_{\mu\nu}(\tau)$. Similarly in Eq.
(\ref{kubo}) we define $\rho^{\rm em}(\omega) = C_{\rm em}\,\rho(\omega)$.
We show in Fig.~\ref{fig:corr} the diagonal components of 
$G_{\mu\nu}(\tau)$. Above $T_c$, correlators measured in ensembles of increasing temperature show little differences between each other, while below $T_c$ their behaviour rapidly changes with $T$.

\section{Maximum Entropy Method}\label{sec:mem}

In this section we  describe the method used to obtain the spectral function.
At large $\omega$, the kernel $K(\tau,\omega)$ in Eq.\ (\ref{spectr})  is
highly suppressed, which allows us to cut off  the
integral  at some
$\omega_\text{max}$. We then discretise the interval $0<\omega<\omega_{\rm max}$ using a stepsize
$\Delta \omega$ with  
 $N_\omega=  \omega_\text{max} / \Delta\omega$. The resulting  equation is 
$ G(\tau_i) =\Delta \omega\, \sum_{j=0}^{N_\omega} \, K_{ij} \rho_j$, where the
 correlator $G(\tau_i)$ is defined over a number
of euclidean points of around $ N_\tau/2\sim O(10-20)$.
 On the other hand, the
discretised spectral function  $\rho_j$ is defined over $N_\omega\sim O(10^3)$
points. This is often referred to as an ill-posed problem.

A search for the solution using standard $\chi^2$ methods would fail, because
the result would not be unique. One way to proceed is to use a physically
motivated Ansatz for the spectral function, with a number of parameters to be
minimized. This path has been followed in Ref.~\cite{fit}.  
  Here we use a bayesian analysis, the Maximum Entropy Method (MEM), which was
first applied to lattice QCD in Ref.\ \cite{mem}. 
  This consists of extracting the most
probable spectral function $\rho(\omega)$, given some prior
knowledge $H$ and the data $D$. This is expressed as a conditional
probability:
\begin{equation}
  P[\rho| D H]=\frac{P[D |\rho H ]P[\rho|H ]}{P[D| H]} \propto \exp(-L + \alpha
S),
\end{equation}
where $L$ is the standard likelihood function and $S$ is the
Shannon-Jaynes entropy, given by:
\begin{equation}\label{default}
 S=\int_0^\infty \frac{d\omega}{2\pi}\left[\rho(\omega)-m(\omega)-
    \rho(\omega)\ln\frac{\rho(\omega)}{m(\omega)}\right],
   \end{equation}
where $m(\omega)=m_0 (b +\omega) \omega$ is the default model, representing our prior
information on $\rho(\omega)$.
Here $m_0$ is an overall normalization and the parameter $b$ is algorithmically crucial to allow a
non-zero conductivity $\sigma$, which is obtained from the 
intercept $\lim_{\omega\rightarrow 0} \rho(\omega)/\omega$, as shown in Eq.\ (\ref{kubo}). 
The result is then obtained by maximising  $P[\rho| D H]$, where we 
use a modification of Bryan's algorithm \cite{Aarts:2007wj} which fixes
kernel instabilities at low $\omega$.

\begin{figure}[t]
\centering
\vskip-2ex
 \includegraphics{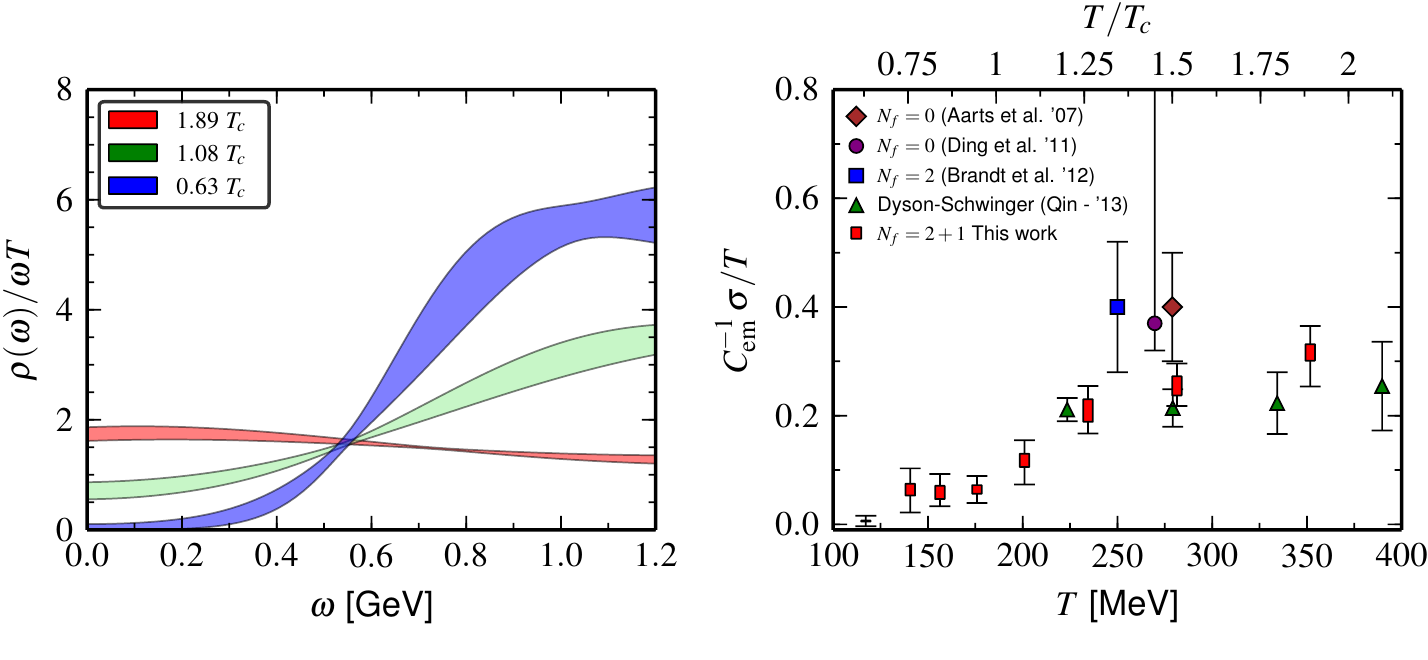}
\caption{Left: spectral functions $\rho(\omega)/\omega T$ at
different $T$'s. The intercept is proportional to $\sigma/T$, see Eq.\
(\protect\ref{kubo}). 
Right: temperature dependence of $C_{\rm em}^{-1}\sigma/T$. The vertical size of the rectangles  
reflects the systematic uncertainty due to changes in the default model, while
the error bars include the statistical jackknife error as well.
  Previous results  \cite{Aarts:2007wj,fit} are
indicated: the $N_f=0$ points are inserted matching the values of
$T/T_c$. The triangles represent a Dyson-Schwinger result \cite{SD}. }
\label{fig:results}
\end{figure}

\section{Results}

In Fig.\ \ref{fig:results} (left) we show the spectral function,
$\rho(\omega)/\omega T$ for three different temperatures. The intercept is
proportional to $\sigma/T$, which is shown in Fig.\ \ref{fig:results} (right)
for temperatures across the deconfining transition. We also compare our $N_f=2$
result to previous findings on the lattice \cite{Aarts:2007wj,fit} and to
results obtained using Dyson-Schwinger equations \cite{SD}.

\begin{figure}[t]
    \centering
\includegraphics[width=\textwidth]{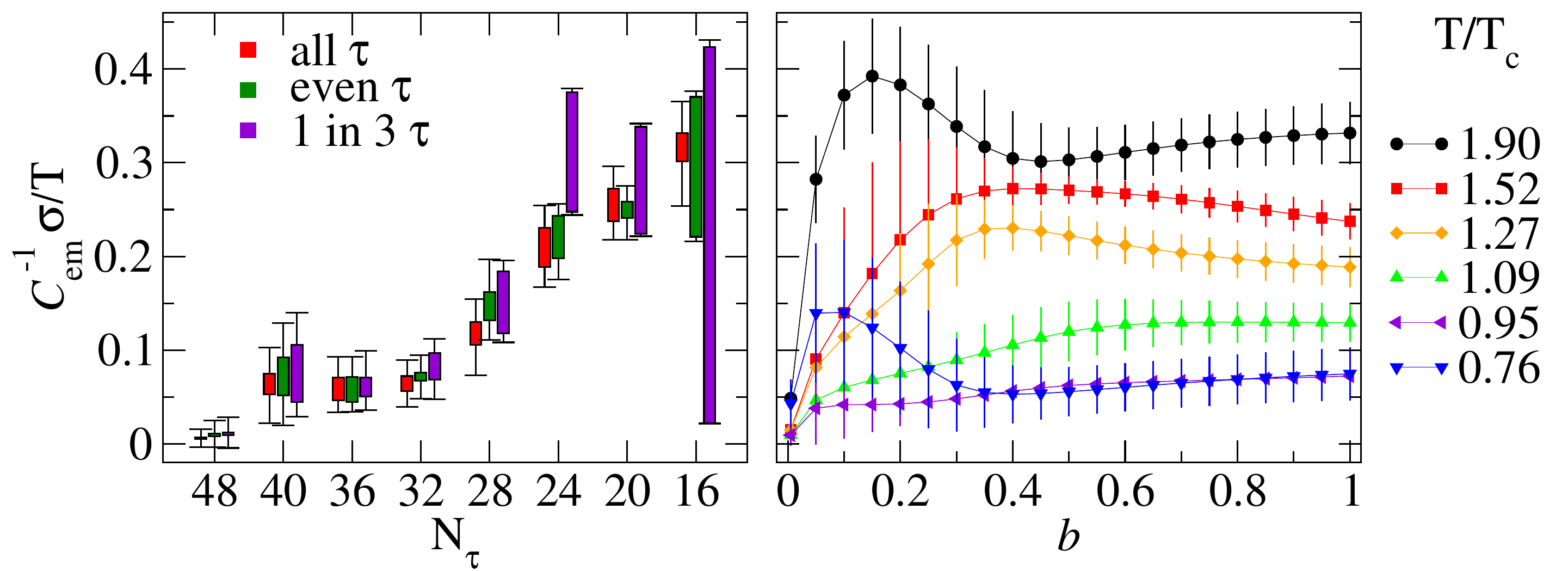}
\caption{
Stability tests. Left: $C_{\rm
em}^{-1}\sigma/T$ when only a
subset of the available time slices is used.
Right: dependence on the parameter $b$
in the default model. Stability is achieved for $b\gtrsim 0.4$.
}
\label{fig:test2}
\end{figure}

We have carried out a series of tests in order to check the 
reliability of the MEM reconstruction of $\rho(\omega)$.
In primis, the choice of default model, aka the specific value of $b$
should not be reflected in the final result for $\sigma$.   In
Fig.~\ref{fig:test2} (right) we explicitly check this by varying $b$ in the MEM analysis, observing stability provided that $b\gtrsim 0.4$.

In Sec.~\ref{sec:simul}  we mentioned the importance of having a high resolution
in the correlator $G(\tau)$ and how this was achieved by the
introduction of the anisotropy. To justify this choice, we run
MEM using only a subset of the available time slices in the
correlator, and check whether the result is stable. This is shown in
Fig.~\ref{fig:test2} (left), where it is clear that for high temperatures, the
anisotropy is crucial to extract a signal for the conductivity.
For colder ensembles the result is instead stable.

\begin{figure}[t]
    \centering
\includegraphics[width=\textwidth]{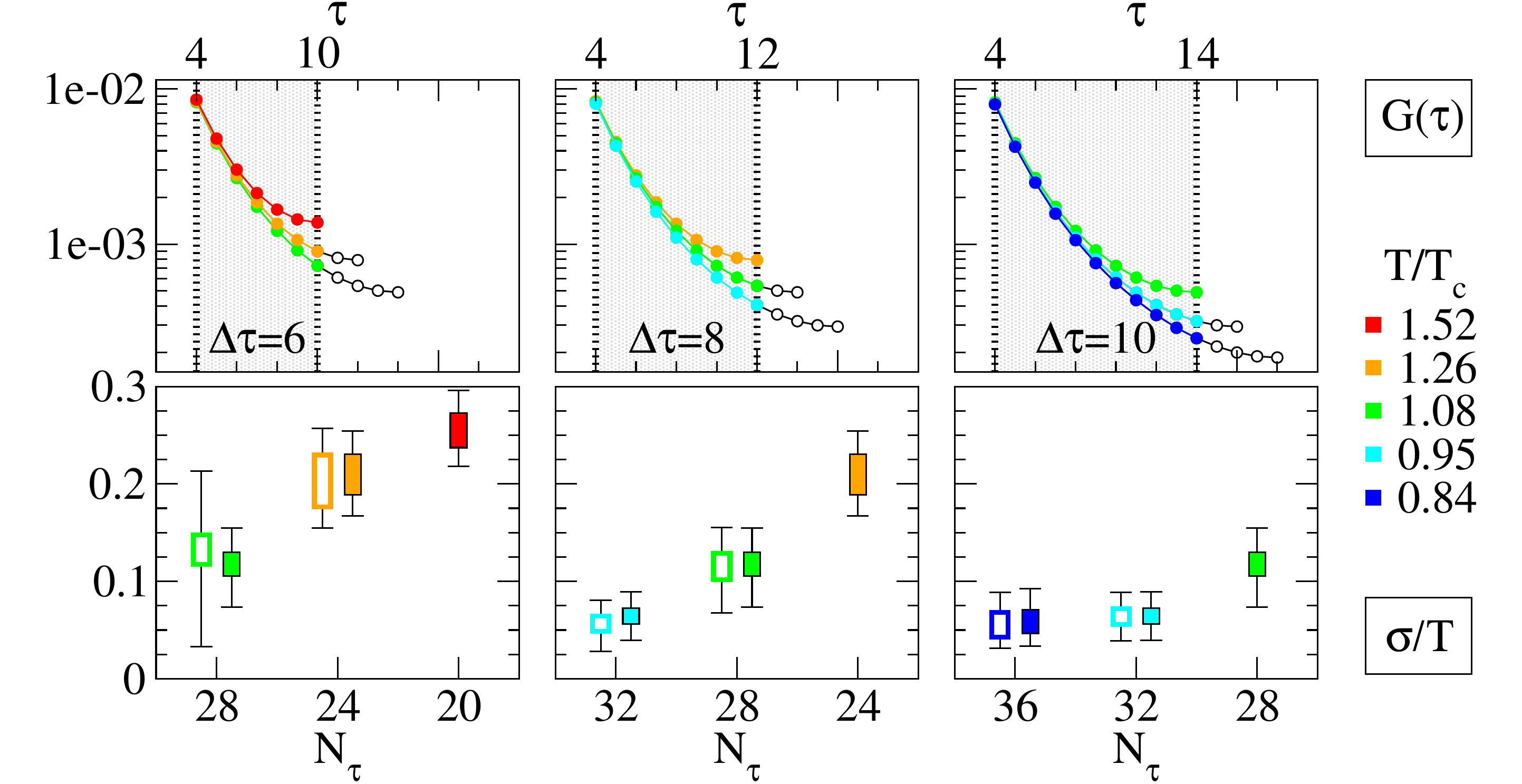}
\caption{Stability tests discarding the last
time slices in the correlator. Top: correlators with used time
slices indicated.
Bottom: corresponding MEM results for $C_{\rm em}^{-1}\sigma/T$.
Open symbols use the restricted time range $\Delta\tau$ shown in the upper
pane, full symbols use the entire time range available.}
\label{fig:test1}
\end{figure}

As also pointed out in Sec.\ \ref{sec:simul}, the number of euclidean
points available in the correlator  will decrease as  the temperature is raised.
One might think that this difference in the extent of $G(\tau)$  is responsible
for the $T$ dependence of the conductivity, rather than it being a genuine thermal
effect. To show that this is not the case, we run the MEM analysis on
correlators at different temperatures, but constrained to have  the same number
of time slices, which is achieved by systematically discarding  the last
points of colder ensembles. A graphical representation of the procedure is
showed in Fig.~\ref{fig:test1}.
We observe excellent stability as the euclidean time range is varied.

 \section{Conclusions}
We have presented the first lattice QCD calculation of the temperature
dependence of the electrical conductivity $\sigma$ divided by $T$, 
using the conserved current and anisotropic lattices \cite{Amato:2013naa}.
We found that $\sigma/T$ increases with temperature.
In the near future we plan to include the contribution from the strange quark
and also evaluate the charge diffusion constant $D$,
combining the results for $\sigma$ with those for the electric charge  susceptibility
\cite{pietro}.

\acknowledgments
This work was supported by STFC, UKQCD and the STFC funded DiRAC Facility,
the Royal Society, the Wolfson Foundation, the Leverhulme Trust and
the European Union Grant Agreement number 238353 (ITN STRONGnet).


\begin{thebibliography}{99}

\bibitem{Amato:2013naa} 
  A.~Amato, G.~Aarts, C.~Allton, P.~Giudice, S.~Hands and J.~-I.~Skullerud,  
    Phys.\ Rev.\ Lett.\  {\bf 111} (2013) 172001,
  [hep-lat/1307.6763].
  
  
\bibitem{Tuchin:2013ie}
  K.~Tuchin,
  arXiv:1301.0099 [hep-ph];
  L.~McLerran and V.~Skokov,
  arXiv:1305.0774 [hep-ph].

\bibitem{effective}
  U.~W.~Heinz and R.~Snellings,
  arXiv:1301.2826 [nucl-th];
  C.~Gale, S.~Jeon and B.~Schenke,
  Int.\ J.\ Mod.\ Phys.\ A {\bf 28}, 1340011 (2013).

\bibitem{Arnold:2000dr} 
  P.~B.~Arnold, G.~D.~Moore and L.~G.~Yaffe,
  JHEP {\bf 0011}, 001 (2000)
  [hep-ph/0010177].
 
\bibitem{review}
  G.~Aarts,
  PoS LAT {\bf 2007} (2007) 001;
 T.~Sch\"afer and D.~Teaney,
  Rept.\ Prog.\ Phys.\  {\bf 72} (2009) 126001;
   H.~B.~Meyer,
  Eur.\ Phys.\ J.\ A {\bf 47} (2011) 86.



  
\bibitem{Kadanoff1963419}
L.~P~Kadanoff and P.~C.~Martin,
Annals of Physics, 24 (1963) 419--469.


\bibitem{anis}
  H.~-W.~Lin {\it et al.}  [Hadron Spectrum Collaboration],
  Phys.\ Rev.\ D {\bf 79} (2009) 034502;
  R.~G.~Edwards, B.~Joo and H.~-W.~Lin,
  Phys.\ Rev.\ D {\bf 78} (2008) 054501.


\bibitem{Aarts:2007wj}
  G.~Aarts, C.~Allton, J.~Foley, S.~Hands and S.~Kim,
  Phys.\ Rev.\ Lett.\  {\bf 99} (2007) 022002.

\bibitem{smear}
  C.~Morningstar and M.~J.~Peardon,
  Phys.\ Rev.\ D {\bf 69} (2004) 054501.

\bibitem{inprep1}
J.-I.~Skullerud et al, in preparation.

\bibitem{gupta}
  S.~Gupta,
  Phys.\ Lett.\ B {\bf 597}, 57 (2004).


\bibitem{fit}
  H.~-T.~Ding, A.~Francis, O.~Kaczmarek, F.~Karsch, E.~Laermann et al,
  Phys.\ Rev.\ D {\bf 83} (2011) 034504;
  B.~Brandt, A.~Francis, H.~B.~Meyer and H.~Wittig,
  JHEP {\bf 1303} (2013) 100.


\bibitem{mem}
  M.~Asakawa, T.~Hatsuda and Y.~Nakahara,
  Prog.\ Part.\ Nucl.\ Phys.\  {\bf 46}, 459 (2001).

\bibitem{SD}
 S.~Qin, 
 arXiv:1307.4587 [nucl-th].

\bibitem{pietro} 
  P.~Giudice, G.~Aarts, C.~Allton, A.~Amato, S.~Hands and J.~-I.~Skullerud,
    arXiv:1309.6253 [hep-lat].
  
\end{thebibliography}
\end{document}